\documentstyle[sprocl,epsfig]{article}

\def\Journal#1#2#3#4{{#1} {\bf #2}, #3 (#4)}
\def\PRD{{\em Phys. Rev.} D}
\def\NPB{{\em Nucl. Phys.} B}
\newcommand{\bqa}{\begin{eqnarray}}
\newcommand{\eqa}{\end{eqnarray}}
\newcommand{\bc}{\begin{center}}
\newcommand{\ec}{\end{center}}
\newcommand{\nn}{\nonumber}
\newcommand{\si}{\sigma}
\newcommand{\ta}{\tau}

\newcommand{\om}{\omega}
\begin{document}

\title{\vspace{-3mm}
TESTING MEM WITH DIQUARK AND THERMAL MESON CORRELATION
FUNCTIONS\footnote{Work supported by the TMR-Network grant ERBFMRX-CT-970122
and the DFG grant Ka 1198/4-1.}$^,$~\footnote{Presented at the Conference on
Strong and Electroweak
Matter (SEWM 2000), Marseille, France, 14-17 June 2000.}}

\author{\vspace{-3mm}I. Wetzorke and F. Karsch}

\address{Fakult\"at f\"ur Physik\\ Universit\"at Bielefeld\\
33615 Bielefeld, Germany}

\maketitle

\vspace{-51mm}
\mbox{} \hfill BI-TP 00/28\\
\mbox{} \hfill August 2000\\
\vspace{40mm}
\abstracts{When applying the maximum entropy method (MEM) to the analysis
of hadron correlation functions in QCD a central issue is to understand to
what extent this method can distinguish bound states, resonances and
continuum contributions to spectral functions. We discuss these issues
by analyzing meson and diquark correlation functions at zero temperature
as well as free quark anti-quark correlators. The latter test the
applicability of MEM to high temperature QCD.}

\section{Principles of MEM}
The {\sc{\bf M}aximum {\bf E}ntropy {\bf M}ethod}
is a common technique in condensed matter physics,
image reconstruction and astronomy~\cite{ja}. In lattice
QCD it has been applied recently to analyze meson correlation
functions at zero temperature~\cite{na}. It could be demonstrated that
this method correctly detects the location of poles in the correlation
function and, moreover, is sensitive to the contribution of higher
excited states to the correlators.

Starting from correlation functions, $D(\tau)$, in Euclidean time which are
calculated on the lattice at a discrete set of time separations,
$\{\tau_k\}_{k=1}^{N_\tau}$, the
application of MEM allows to extract the {\it most probable}
spectral function without any assumptions about the spectral shape.
This is an important key to access real time correlation functions
and other dynamical quantities through lattice calculations. Of course,
the essential bottleneck is that one has to specify what is meant by
the {\it most probable spectral function}:
MEM is based on {\bf Bayes theorem of conditional probability},
which yields the most probable result for the spectral function
$A(\omega)$ given the dataset $D(\tau)$ and all prior knowledge $H$
(e.g. positivity of $A(\omega)$)~\cite{ja,br}:
\bqa\!\!
P[A|DH] \sim {P[D|AH]} {P[A|H]}
\mbox{~~with~}
\begin{array}{ccl}
  {P[D|AH]} &\sim& \exp(-{L})\\
  {P[A|H]} &\sim& \exp({ \alpha S})
\end{array}
\label{bayes}
\eqa
The {\bf Likelihood function}, $L$, is chosen as the usual $\chi^2$
distribution
\bqa
L = {1\over2} \chi^2 = {1\over2} \sum
(F(\tau_i)-D(\tau_i)) C_{ij}^{-1} (F(\tau_j)-D(\tau_j))\nn
\eqa
with $C_{ij}$ denoting the symmetric covariance matrix constructed
from the data sample and $F(\tau) = \int d\omega K(\tau,\omega) A(\omega)$
is the {\it fit function} in terms of a predefined kernel
$K(\tau,\omega)$ and the spectral function
$A(\omega)\equiv \rho(\omega) \; \omega^2$.
For an unnormalized probability function $A(\omega)$ we can parametrize
the {\bf entropy},
\bqa
S = \int d\omega [A(\omega) - m(\omega) - A(\omega)
\log(\frac{A(\omega)}{m(\omega)})]~~,\nn
\eqa
where $m(\omega) = m_0 \; \omega^2$ is the {\it default model},
i.e. the initial ansatz for $A(\omega)$.
The default model incorporates our
knowledge about the short distance behaviour of meson correlation
functions~\cite{na}, which in leading order perturbation theory are
proportional to $\omega^2$. The real and positive factor
${\bf \alpha}$ controls the relative weight between the entropy (default
model) and Likelihood function (data) appearing in Eq.~\ref{bayes}.

The most probable spectral function $\hat{A}_\alpha(\omega)$ is then
obtained by maximizing $ Q \equiv {\alpha S} - {L}$ for
given $\alpha$ and the final spectral function
$\bar{A}(\omega)$ is determined from a weighted average over $\alpha$:
\bqa
\bar{A}(\omega) = \int {\cal D}A \;d\alpha \;A(\omega) P[A|DH]
P[\alpha|D] \simeq \int d\alpha \; {\hat A}_\alpha(\omega)
P[\alpha|D]~~. \nn
\eqa
Usually it turns out that the weight factor $P[\alpha|D]$ is sharply
peaked around a unique value ${\hat \alpha}$.

In the following we will apply this general framework to analyze
zero-momentum meson and diquark correlation functions in Euclidean time,
$D(\tau)$. The correlators are
calculated in quenched QCD on lattices with temporal extent $N_\tau$.
They can be expressed in terms of the
spectral function $A(\omega)$,
\bqa
D(\tau) = \int d^3x \;\langle{\cal O}^\dag(\tau,\vec{x})
{\cal O}(0,\vec{0})\rangle
= \int d\omega \;K(\tau,\omega) A(\omega)~~.\nn
\eqa
Here the kernel $K(\tau,\omega)$ is taken to be proportional to the
Fourier transform of a free boson propagator,\\[-5mm]
\bqa
K(\tau,\omega) = e^{-\tau\omega}/(1 - e^{-N_\tau\omega})~~.\nn
\eqa
The computational intensive part of MEM is the maximization of
$Q = \alpha S - L$ in the functional space of $A(\omega)$, for
which one typically uses a few hundred degrees of freedom.
A {\bf S}ingular {\bf V}alue {\bf D}ecomposition (SVD) of the kernel
$K(\tau, \omega)$ was performed to reduce the parameter space.

\section{Details of the Simulation}
For our tests of MEM we used data from a previous spectrum
calculation~\cite{he} in quenched QCD
with Wilson fermions on lattices of the size $16^3 \times 30$ and $16^3
\times 32$. In order to study also diquark correlation functions
Landau gauge fixing has been performed on the larger lattice. In the
gluon sector we use the Symanzik improved (1,2) action, which
eliminates ${\cal O}(a^2)$ cut-off effects. In the fermion sector we
use the ${\cal O}(a)$ improved clover action with a tree level Clover
coefficient. Our analysis is based on
73 gauge field configurations generated at $\beta$ = 4.1, which corresponds
to a lattice spacing $a^{-1}\simeq 1.1$~GeV. The fermion matrix has been
inverted at eight different values of the hopping parameter $\kappa$.
These cover a range of quark masses between $\sim$30~MeV and $\sim$250~MeV.
On each configuration we use four random source vectors at different
lattice sites. We thus have a dataset consisting of 292 quark propagators
for each quark mass value. Correlation functions were obtained for
pi- and rho-meson, as well as for diquark states in the color anti-triplet
representation with an attractive q-q interaction and the repulsive color
sextet channel~\cite{he}.

\section{Meson Spectral Functions}
The spectral functions in pseudo-scalar and vector meson quantum number
channels obtained from our analysis using MEM are shown in Fig.~1.
In addition to the dominant ground state peaks
(pi- and rho-mesons) in the low-energy region, the meson spectral functions
show signs of excited states and a broad continuum-like structure at high
energies up to the momentum cut-off on the
lattice $\omega_{\rm max}=\pi/a \sim 3.5$~GeV.
We generally find that better statistics and larger $N_\ta$ result in higher
and narrower peaks in the spectral functions which indicates that these
correspond to $\delta$-function like singularities, {\it i.e.} poles
in the propagator.
We also note that in general the ground state contribution is more
dominant for lighter quark masses. This is clearly seen in the
pseudo-scalar spectral function. On the other hand, the broadening and
drop of the low mass contributions seen for the lightest quark masses in
the vector spectral function can be addressed to insufficient statistics
for this correlator. This is also apparent from conventional
exponential fits, which in these cases lead to large errors on the
lightest vector meson masses.

Comparing the MEM results with standard two-exponential fits of the
correlation functions we find very good agreement. The point of the
vanishing pion mass is $\kappa_c$ = 0.14922(1), which
perfectly coincides with the value 0.14923(2) obtained from a
conventional exponential
fit. Extrapolating the mass of the rho meson to the chiral limit we obtain
$m_\rho$ = 0.56(3) in lattice units, which should be compared with
$m_{\rho}$(exp-fit) = 0.58(2).

\begin{figure}
\hspace*{-3mm}
\epsfig{file=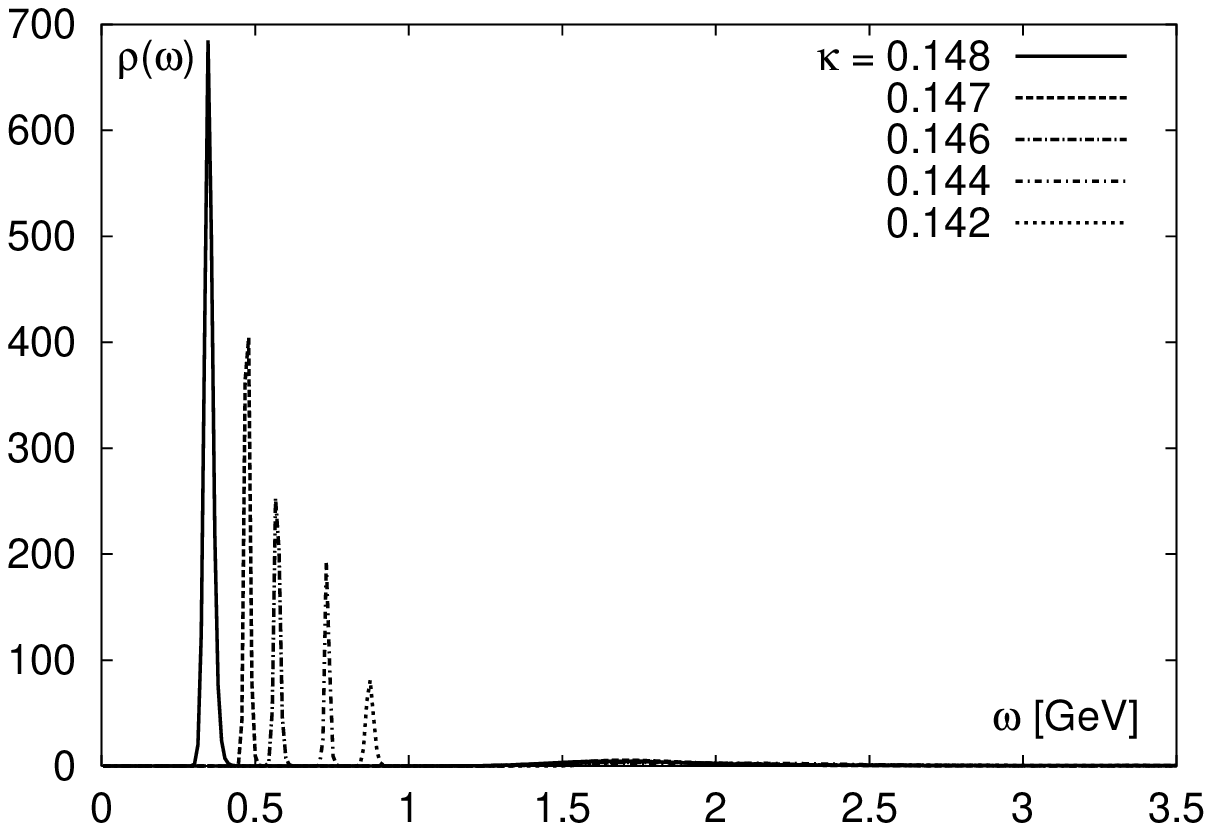,width=6cm}
\hfill
\epsfig{file=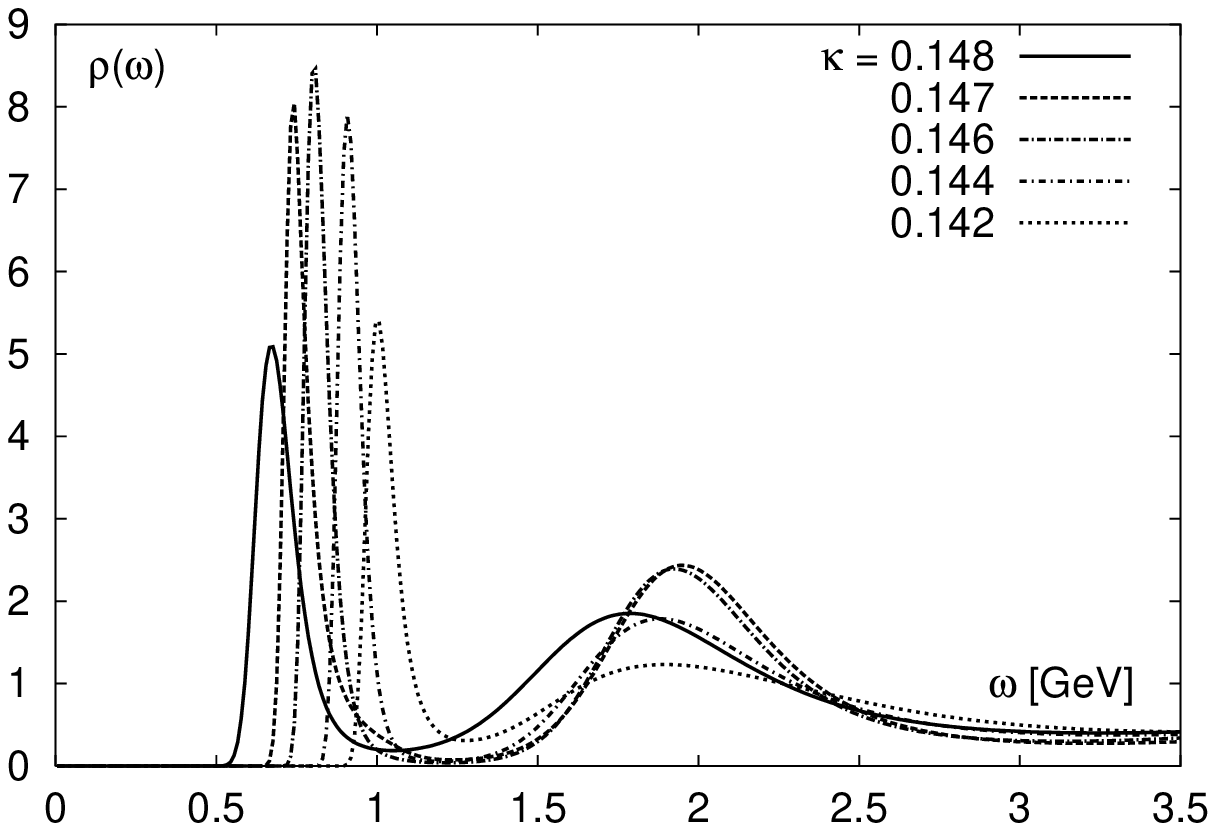,width=6cm}
\caption{Zero temperature spectral functions for pion (left) and rho meson
  (right) for different values of the bare quark mass, (note the different
  $\rho(\omega)$ scale).}
\vspace*{-3mm}
\end{figure}

It was shown by the UKQCD group~\cite{la} that smeared operators and gauge
fields provide a better overlap with the ground state. We applied the fuzzing
technique for spatially extended operators with a radius $R$ = 0.7 fm.
The spectral functions of the fuzzed mesons show only
the ground state peak. They thus uncover unambiguously that the application
of the fuzzing technique eliminates the excited states almost completely.

\section{Diquark Spectral Functions}
We previously had studied diquark correlation functions in Landau
gauge~\cite{he}. In particular the correlators for color sextet states
seemed to be strongly influenced by higher excited states and it was
unclear whether a bound state exists at all in these quantum number
channels.
These difficulties are also show up in our spectral analysis.

Spectral functions similar to those for mesons were obtained
for the color anti-triplet diquark states (Fig.~2(a)).
In particular the $(\bar{3}0\bar{3})$ diquark shows a pronounced
ground state peak. However, already in the case of the $(61\bar{3})$
state this ground state peak broadens and becomes comparable in
magnitude to the broad continuum structure found at higher energies.
This broad continuum becomes even more dominant in the color sextet
channels $(\bar{3}16)$ and $(606)$ shown in Fig.~2(b).
In fact, in the case of color sextet states we find that the spectral
function is strongly dependent on the large momentum cut-off
$\omega_{\rm max}$ used in our analysis. An analysis at smaller lattice
spacings thus is needed to better understand the spectral functions
in these quantum number channels.

\begin{figure}[ht]
\hspace{-3mm}
\epsfig{file=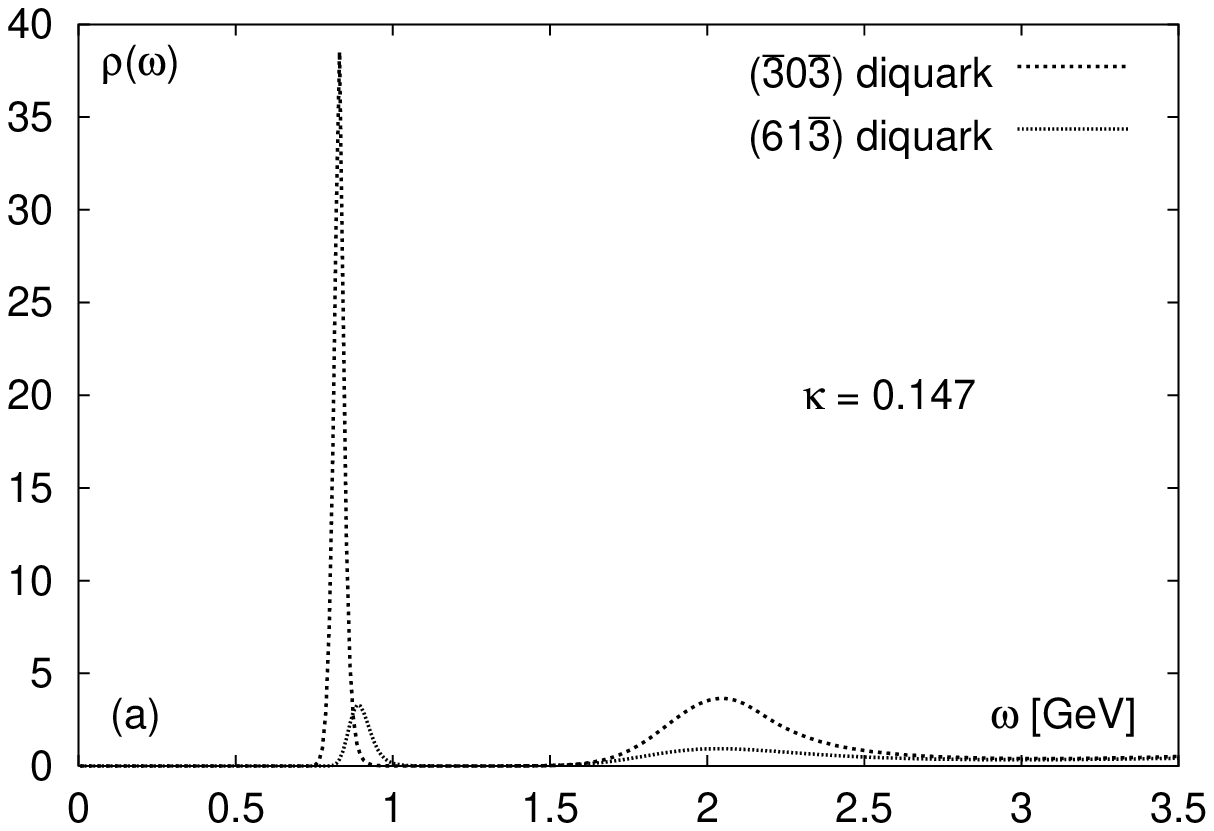,width=6cm}
\hfill
\epsfig{file=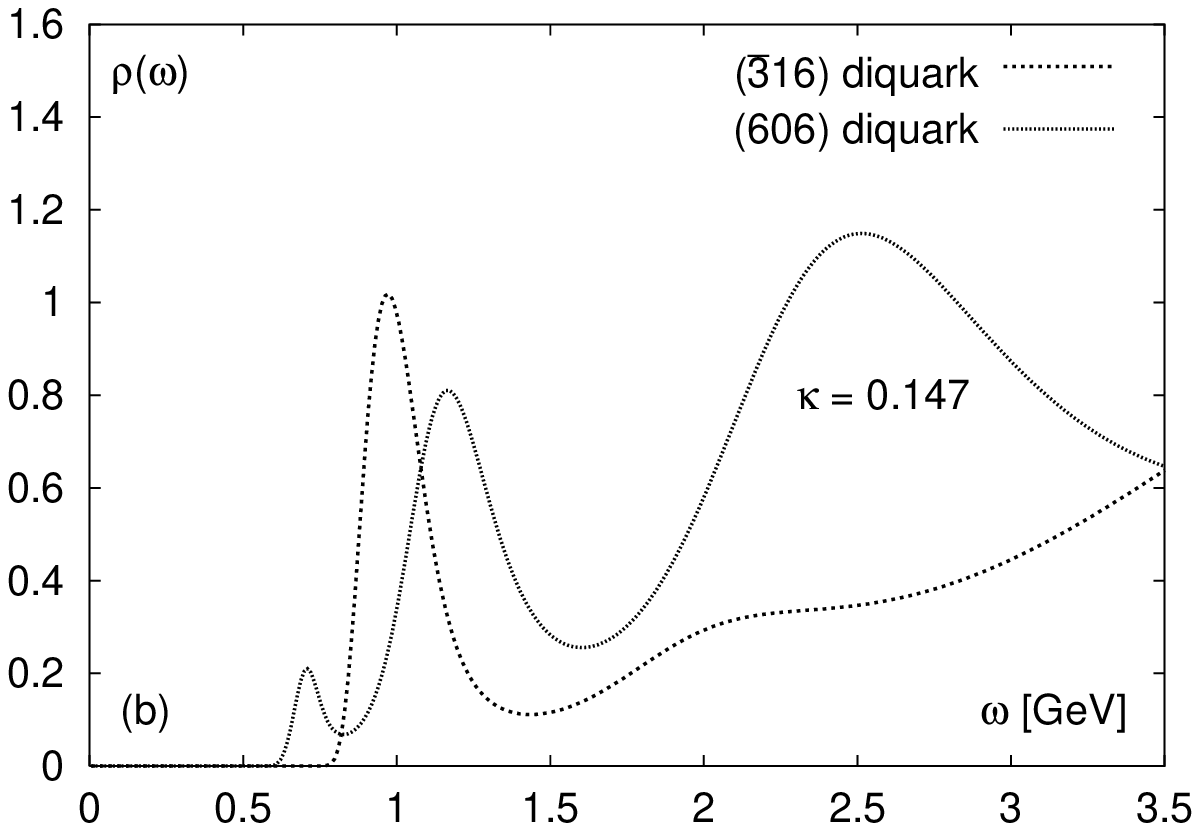,width=6cm}
\caption{Spectral functions for color anti-triplet (a)
  and sextet diquark states (b).}
\vspace*{-3mm}
\end{figure}

The MEM results for the color anti-triplet diquarks are in good
agreement with the masses extracted from two-exponential fits (see Tab.~1).
For the color sextet diquarks we observe a deviation in
comparison with the 2-mass fits. The mass of the $\bar{3}16$
diquark has about the same value in the chiral limit, whereas
the $606$ diquark mass obviously is not well represented by a unique
mass. This favours
the interpretation of color sextet states being unbound, in accordance with
the expected repulsive q-q interaction in this channel~\cite{he}.
\begin{table}[ht]
\renewcommand{\arraystretch}{1.1}
\bc
\vspace{-1mm}
\begin{tabular}{|l|cccc|}
\hline
Diquark state &
($\bar{3}0\bar{3}$) & ($61\bar{3}$) & ($\bar{3}16$) & ($606$) \\
\hline
$ma_{(FSC)}$: MEM result & 0.60(2) & 0.70(3) & 0.74(9)\hspace*{2mm}  & ---   \\
$ma_{(FSC)}$: 2-exp. fit & 0.62(2) & 0.73(4) & 0.77(17) & 0.50(15)\\
\hline
\end{tabular}
\ec
\caption{Diquark masses in the chiral limit obtained with MEM and
two-exponential fits. Our notation for quantum numbers of the diquark
states is ({\bf F}lavor, {\bf S}pin, {\bf C}olor)-representation.}
\vspace*{-5mm}
\end{table}

\section{Thermal Spectral Functions}
At non-zero temperature meson correlation functions are periodic
and restricted to the Euclidean time interval $[0,1/T]$,
\bqa
\!D(\tau)\!=\!\int_0^\infty\!d\om A(\om)
\frac{\cosh(\om(\ta - 1/2T))}{\sinh(\om/2T)}~~.\nn
\eqa
In the high temperature limit the meson spectral functions are
expected to approach those of freely propagating quark anti-quark pairs.
To leading order this is described by the free spectral function,
{\it i.e.}
$\!A(\om)\!=\!\frac{N_c}{4\pi^2}\;\om^2 \tanh(\om/4T)$ in the scalar
channel. We want to test
here whether this behaviour can be reproduced on lattices with
finite temporal extent $N_\ta$. We use the continuum expression for
$D(\tau)$ and evaluate it at a discrete set of
Euclidean times, $\ta_k= k/N_\ta$, with $k=1,~2,..., N_\ta-1$. This is
shown in Fig.~3(a). The reconstructed spectral functions in Fig.~3(b)
were obtained with MEM,
adding Gaussian noise with the variance $\si(\ta) = b \;
D(\ta)/\ta$ to the exact results.
\begin{figure}[ht]
\hspace*{-3mm}
\epsfig{file=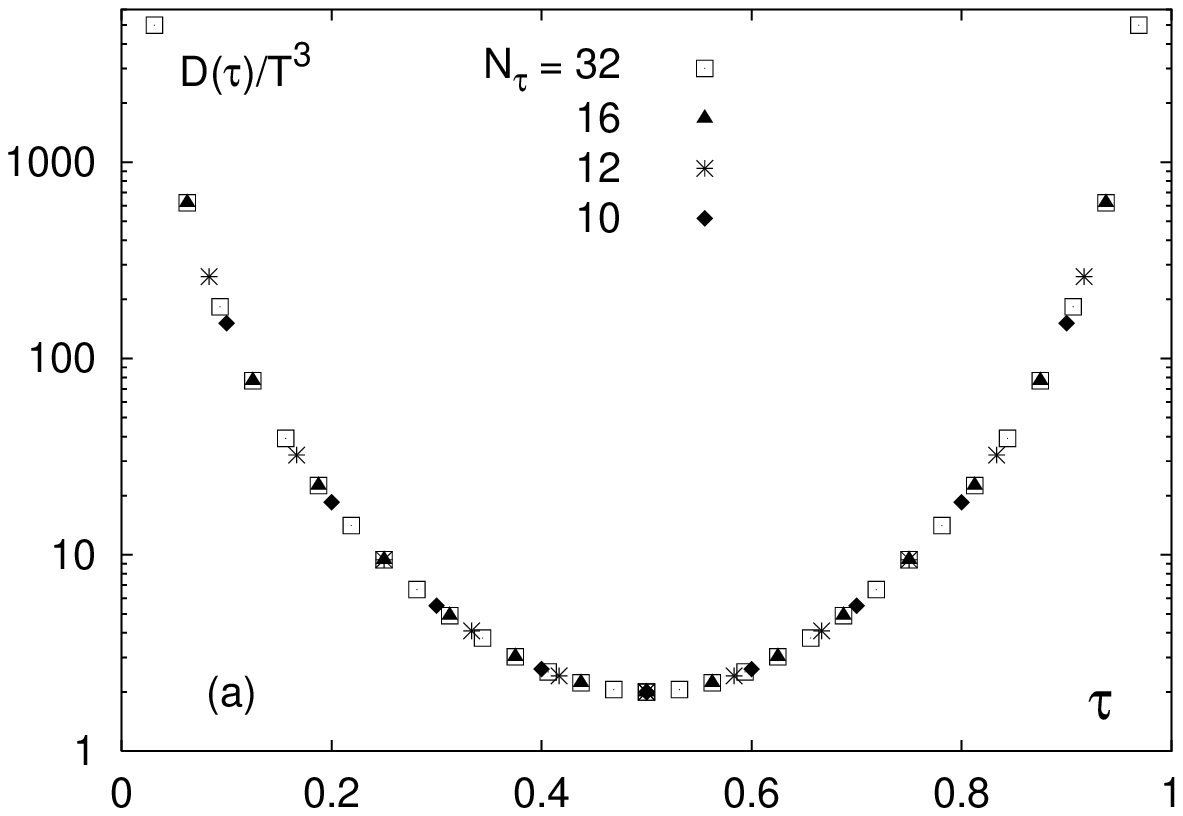,width=6cm}
\hfill
\epsfig{file=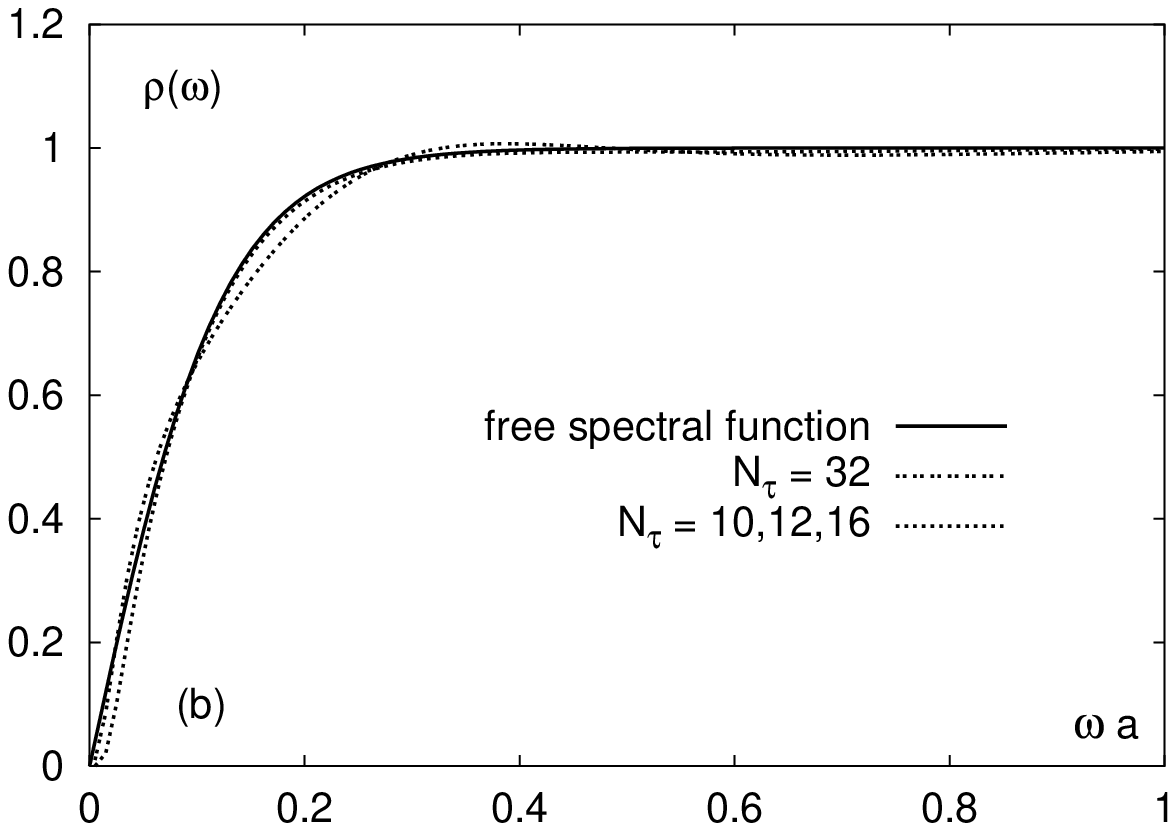,width=6cm}
\caption{Free thermal meson correlation function at Euclidean times
$\tau \equiv \tau_k=k/N_\tau$ for different $N_\tau$ (a) and the
reconstructed spectral
function (b). For the reconstruction shown in (b) different subsets of the
data shown in (a) have been used as indicated. The noise level on the
data has been set to $b=0.01$.}
\vspace*{-5mm}
\end{figure}

\vspace*{0.1mm}
We note that one can reproduce the shape of the spectral function quite
well already from correlation functions calculated at $N_\ta=16$
points. However, twice as many data points are needed for a good
quantitative description of $\rho(\omega)$.
One thus may expect that the reconstruction of the continuum part
of thermal correlation functions based on simulations on lattices
with temporal extent $N_\tau$ will require information on the
correlation functions at ${\cal O} (30)$ points. This may either be
achieved by performing calculations on large temporal lattices or
by combining information from lattices with different temporal
extent but fixed temperature.
\vspace*{-1mm}
\section{Conclusions}
We find that the application of MEM to the analysis of lattice
correlation functions does yield useful additional information
to that of conventional exponential fits. It can lead to quantitative
results on pole and continuum contributions to spectral functions.

\end{document}